\documentclass[prl,twocolumn,showpacs,floatfix,aps]{revtex4}
\usepackage{graphicx}

\begin{document}
\preprint{cond-mat/0000000}

\title{Superconducting gap and electron-phonon interaction in MgB$_{2}$ thin
film studied by point contacts\footnote{Presented at the NATO ARW
"New Trends in Superconductivity", Yalta, Ukraine, 16-20 Sept.
2001.}}

\author{N. L. Bobrov, P. N. Chubov, Yu. G. Naidyuk\footnote{Corresponding
author, e-mail: naidyuk@ilt.kharkov.ua}, L.V. Tyutrina, and I. K.
Yanson}

\affiliation{B.Verkin Institute for Low Temperature Physics and
Engineering, National Academy  of Sciences of Ukraine, 47 Lenin
Ave., 61103,  Kharkiv, Ukraine}

\author{W. N. Kang, Hyeong-Jin Kim, Eun-Mi Choi, and Sung-Ik Lee}

\affiliation{National Creative Research Initiative Center for
Superconductivity, Department of Physics, Pohang University of
Science and Technology, Pohang 790-784, South Korea}

\date{\today}

\begin{abstract}
The superconducting order parameter and electron-phonon
interaction (EPI) in a MgB$_{2}$ c-axis oriented thin film are
investigated by point contacts (PCs). Two superconducting gaps
$\Delta_{Small}\approx 2.6$\,meV and $\Delta_{Large}\approx 7.4$
meV have been established, which fairly well correspond to the
theoretical gap ratio 1:3. The reconstructed PC EPI function along
c-axis has above 30 meV similarities with the phonon DOS, while a
pronounced maximum around 20\,meV is resolved pointing out that
the low frequency phonon mode can be of importance for pairing. We
have estimated the Fermi velocity in MgB$_2$ about 5$\times
10^7$\,cm/s.

\end{abstract}

\maketitle

\section{Introduction}

The compound MgB$_2$ with graphite-like planes of boron atoms
attracts much attention as superconductor with at the present the
highest T$_c\simeq$40\,K  \cite{Nagam} for a binary systems. The
observation of boron isotope effect \cite{Budko,Hinks} and
examination of electron-phonon coupling
\cite{Kortus,Kong,Bohnen,AP,Liu,Yildirim,Osborn} in MgB$_2$ are in
accordance with the expectations for conventional BCS
superconductivity mediated by electron-phonon interaction (EPI).
Investigations of the order parameter  by tunneling
\cite{Karapet,Rubio,Sharoni,Chen,Giubileo,Giubileo2,Badr,Zasad}
and point-contact technique
\cite{Zasad,Deuts,Plec01,Szabo,Laube,Gonelli,Bugos} confirm that
MgB$_{2}$ is most likely an $s$-wave superconductor. In all cases
the spectra  show unambiguous features of an energy gap $\Delta $
in the density of states (DOS), albeit the results are
controversial on the gap width. Values of $\Delta $ ranging from
1.5 to 8\,meV have been reported, pointing out the possibility of
an anisotropic or distributed (nonhomogeneous) energy gap or even
multiple gaps. The latter scenario has been recently recalled by
Liu et al. \cite{Liu} for MgB$_2$. The Fermi surface consisting
from nearly cylindrical hole sheets arising from quasi-2D boron
bands and three dimensional tubular network \cite{Kortus} was
considered. The different character of the sheets raises the
possibility that each has a distinct gap. The existence of two
different energy gaps with the ratio approximately 1:3 being
respectively smaller (for the 3D gap) and larger (for the 2D gap)
than the standard weak coupling BCS value $\Delta $=1.76k$_{\rm
B}$T$_c\simeq$6\,meV was predicted in \cite{Liu}.

Determination of the Eliashberg function $\alpha^2$F for
superconducting systems provides a consistency check of the
phonon-mediated pairing mechanism. In
\cite{Kong,Bohnen,AP,Liu,Yildirim,Osborn} the phonon dispersion
and DOS, the EPI function $\alpha^2$F and electron-phonon coupling
constant $\lambda$, T$_c$ and isotope effect for MgB$_2$ have been
calculated using different methods and approximation. Estimation
of the electron-phonon-coupling strength in MgB$_2$
\cite{Kortus,Kong,Bohnen,AP,Liu,Yildirim} yields $\lambda \simeq
$1. Established  by different authors, the phonon DOS and the EPI
function have both similarities and differences. The common
feature for presented in \cite{Bohnen,Yildirim,Osborn,Clemen}
phonon DOS is the maximum energy of about 100 meV, absence of
visible phonon peaks below 30 meV and rich structure with a number
of peaks in between 30-100 meV. The main feature of the calculated
EPI function is a dominant maximum between 60 and 75 meV arising
from in-plain boron E$_{2g}$ phonon mode. The precise energy
position and shape of the maximum depend on the method of
calculation \cite{Kong,Bohnen,Liu}. The transport EPI function, as
shown in \cite{Kong}, mimics the EPI one, although, as mentioned
in \cite{Liu}, interband anisotropy reduces the transport coupling
constant from 1 to 0.5. From both the DC resistivity and optical
conductivity measurements \cite{Tu}, even a smaller value of
$\lambda_{tr}$ = 0.13 is derived.

However, some neutron inelastic scattering measurements with
higher resolution \cite{Mura,Sato} revealed a clear maxima in
generalized phonon DOS below 30 meV, namely at 16 and 24 meV in
\cite{Mura} or at 17.5 meV in \cite{Sato}. Therefore it is an open
question whether the only high frequency phonons are responsible
for thermodynamic, transport and especially for fascinating SC
properties of MgB$_2$.

It has been also suggested in \cite{Manske} that the "multi-gap"
structure observed in the tunneling spectra \cite{Giubileo} can be
explained by considering of a low-frequency phonon mode at 17.5
meV, which is revealed in the inelastic neutron scattering
experiment \cite{Sato}. Soft bosonic mode was also exploiting in
\cite{Shulga} to describe an upper critical field behavior of
MgB$_2$ within a multi-band Eliashberg model.

Measuring of a nonlinear conductivity of point contacts (PC)
between two metals allows in a direct way recovering the EPI
function $\alpha^2$F \cite{Yanson}, as well as investigating the
superconducting (SC) gap \cite{BTK82}. The goal of this report is
simultaneous study of peculiarities of EPI along with SC order
parameter of MgB$_2$ by PC method in order to clarify the
mentioned issues, namely, the multigap structure of the order
parameter and the peculiarities of EPI function, trying to find
their correlation.

\section{Experimental and results}

We have used the high-quality c-axis oriented MgB$_2$ thin films
grown by a pulsed laser deposition technique \cite{Lee01}. The
films were grown on (1 $\overline{1}$ 0 2) Al$_2$O$_3$ substrates.
The typical film thickness was 0.4 $\mu$m. The resistivity of the
film exhibits a sharp transition at 39 K with a width of $\sim$
0.2 K from 90\% to 10\% of the normal state resistivity
\cite{Lee02}. The residual resistivity $\rho_0$ at 40 K is $\sim$
6 $\mu\Omega$\,cm \cite{rho} and RRR=2.3.

Different contacts were established in situ at liquid $^{4}$He
temperatures  by touching  as prepared surface of the MgB$_2$ film
by the sharpened edge of an Ag counterelectrode, which were
cleaned by chemical polishing. By this means the current flow
through PCs is preferably along the c-axis. The experimental cell
with the samples holder was immersed directly in liquid $^{4}$He
to ensure good thermal coupling. Both the  differential resistance
d$V/$d$I$ and d$^2V$/d$I^2(V)$ vs $V$ were registered using a
standard lock-in technique. The resistance $R_N$  at $V\gg\Delta$
of investigated contacts ranges from 10 to 1000 $\Omega$ at
4.2\,K.

\subsection{Superconducting gap}

According to the Blonder-Tinkham-Klapwijk theory \cite{BTK82} of
conductivity of N-c-S metallic junctions (here N is normal metal,
c is constriction and S is superconductor) a maximum at zero-bias
voltage and a double-minimum structure around $V\simeq \pm\Delta
/$e in the d$V/$d$I$ curves manifest the Andreev reflection
process with a finite barrier strength parameter $Z$. Thus the
positions of the minima roughly reflect the SC gap value. This
follows from the equations for the current-voltage characteristics
\begin{eqnarray}\label{BTKeq}
I(V) &\sim &\int_{-\infty }^{\infty}
T(\epsilon)\left(f(\epsilon-{\rm e}V)-f(\epsilon)\right) {\rm
d}\epsilon,  \\ T(\epsilon) &=&\frac{2\Delta
^2}{\epsilon^2+(\Delta
^2-\epsilon^2)(2Z^2+1)^2},~~~|\epsilon|<\Delta \nonumber \\
T(\epsilon)
&=&\frac{2|\epsilon|}{|\epsilon|+\sqrt{\epsilon^2-\Delta
^2}(2Z^2+1)},~~~ |\epsilon|>\Delta ~, \nonumber
\end{eqnarray}
where f($\epsilon$) is the Fermi distribution function. The
broadening of the quasiparticle DOS in the superconductor can be
taken into account if $\epsilon$ will be replaced by
$\epsilon-i\Gamma$ in Eq. (1). We used Eq.(1) to fit the measured
d$V/$d$I$ curves of PCs and extract SC gap.

\begin{figure}
\includegraphics[width=\columnwidth,angle=0]{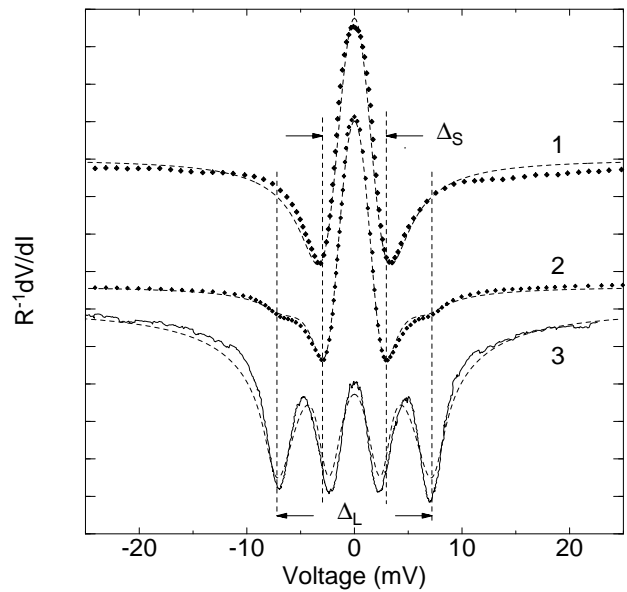}
\caption[] {Reduced differential resistance $R_N^{-1}$d$V$/d$I$
vs. $V$ measured for three  MgB$_2$-Ag contacts with one gap
minima (curve 1, $R_N$=350 $\Omega$) and two gaps (curve 2,3,
$R_N$=20 and $\sim$25$\Omega$) structures. Symbol traces and
bottom solid line are the experimental curves at T = 4.2 K. Thin
dashed lines are theoretical dependence calculated using Eq.(1)
with one gap 2.7 meV (curve 1, Z=0.9, $\Gamma$=1.2 meV) and two
gaps: 7.1 and 2.7  meV (curve 2, Z=0.9, $\Gamma$=0.26 meV), 7.4
and 2.6 meV (curve 3, Z=0.64 and 0.38, $\Gamma\simeq0$). The gap
weight factor ratio  W$_{Small}$/W$_{Large}$ is 10 and 1.5 for
curves 2 and 3, respectively. The y-axis one division corresponds
to 0.05 for curve 1 and 0.1 for curve 2.} \label{fig1}
\end{figure}

The d$V/$d$I$ curves taken at 4.2\,K $\ll T_c$ show usually
structure, which contains symmetric with respect to zero bias
minima at about $\pm$2-3 mV with zero-bias maximum at $V$=0 (see
Fig.~\ref{fig1}, curve 1), related to the SC gap $\Delta$. Often
the d$V$/d$I$ vs. $V$ curves show clear two gap structure with
shallow features corresponding to a larger gap, as shown by curve
2 in Fig.~\ref{fig1}. Among the many d$V$/d$I$ curves (about 100),
there was a curve 3 (Fig.~\ref{fig1}) exhibiting almost equal
features corresponding to the both gaps with minima at about
$\pm$7 and $\pm$2.3\,mV. These values are close to the maxima at 7
and 1.5-2\,mV of the minima distribution in d$V$/d$I$ given in
\cite{Laube}. Fitting of curve 3 by Eq.\,(1) yields the gap value
7.4 and 2.6\,meV, which is consistent with PC measurements 7 and
2.8\,meV in \cite{Szabo} or 6.2$\pm$0.7 and 2.3$\pm$0.3\,meV in
\cite{Bugos} and with single small gap 2.6 meV \cite{Gonelli}. In
our case, the ratio 2.85 of the larger gap to the lower one is
close  to the theoretical value 3:1 \cite{Liu}.

\subsection{Phonons features}

The voltage $V$ applied to the ballistic contact defines the
energy scale e$V$ for scattering processes. EPI results in
backflow scattering of energized electrons so that some of them
are reflected by creating of phonons and do not contribute to the
current. This leads to a nonlinear $I-V$ characteristic and, as
shown by the theory of Kulik, Omelyanchouk and Shekhter
\cite{KOS}, second derivative d$^2V$/d$I^2(V)$ of the $I-V$ curve
is proportional at low temperatures to $\alpha_{PC}^2\,F(\omega)$:
\begin{equation}
\label{pcs} \frac{{\rm d}^2V}{{\rm d}I^2}\propto R^{\rm
-1}\frac{{\rm d}R}{{\rm d}V}= \frac{8\,ed}{3\,\hbar v_{\rm
F}}\alpha_{PC}^2(\epsilon)\,F(\epsilon),
\end{equation}
where $\alpha_{PC}$, roughly speaking, describes the strength of
the electron interaction with one or another phonon branch and
takes into account the kinematic restriction of electron
scattering processes in the point contact by factor
$1/2(1-\theta/\tan\theta)$ (for transport and Eliashberg EPI
functions the corresponding factors are $(1-\cos\theta)$ and 1,
respectively), where $\theta$ is the angle between initial and
final momenta of scattered electrons. It should be mentioned that
in PC the large angle $\theta \to \pi$ scattering
(back-scattering) processes of electrons are strongly underlined.

The PC resistance is determined by a sum of ballistic Sharvin and
diffusive Maxwell term  according to the  simple formula derived
by Wexler \cite{Wexler}:
\begin{equation}
\label{Rwex} R_{\rm PC}(T) \simeq  \frac {16 \rho l}{3\pi d^2} +
\frac{\rho (T)}{d},
\end{equation}
which is commonly used to  estimate the PC  diameter $d$. Here
$\rho l = p_{\rm F}/n$e$^2$, where $p_{\rm F}$ is the Fermi
momentum, $n$ is the density of charge carriers. The latter for
MgB$_2$ is estimated in $n\simeq 6.7\times 10^{22}$
\cite{Canfield}, which results in $\rho l\simeq 2\times
10^{-12}\Omega\cdot $cm$^2$ using $v_{\rm F}\simeq 5\times
10^{7}$cm/s \cite{Kortus}. Hence, the upper limit for elastic mean
free path for our film is about 3 nm. In this case, using Eq.\,
(3), we can find that condition $d<l$ is fulfilled for PC with
$R>40\,\Omega $ or for lower resistance supposing multiple
contacts in parallel.

The measurements of d$^2V$/d$I^2(V)$ dependencies to recover
spectral EPI function reveal a wide variety of curves.
Nevertheless, we were able to select similar d$^2V$/d$I^2(V)$
characteristics for PC with different resistance (see Fig.2). The
common features are the reproducibility of position of the main
maxima placed at about 40, 60 and 80-90 meV and lack of spectral
features above 100 meV. There is a correlation in the peak
position in d$^2V$/d$I^2(V)$ and calculated in \cite{Liu} EPI
function (Fig.\,2).

\begin{figure}

\includegraphics[width=\columnwidth]{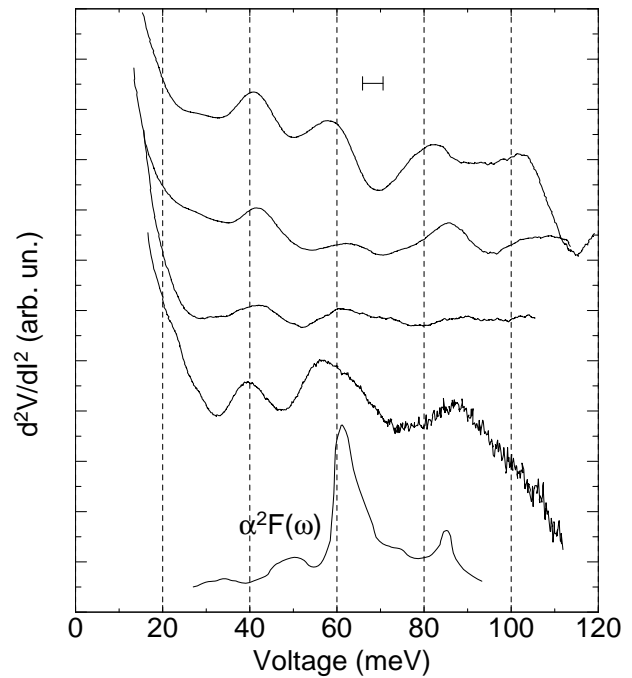}
\caption[] {Second derivative d$^2V$/d$I^2(V)$ averaged for
opposite polarities for 4 MgB$_2$-Ag contacts with different
resistance (R = 43, 36, 111 and 45 $\Omega$ from top curve to
bottom one) at T=4.2 K with clear maxima in the range of phonons
above 30 meV. Bottom curve is calculated in \cite{Liu} smoothed
with experimental resolution (the segment above) EPI function.
Signal increase in experimental curves below 30 meV is due to the
SC gap structure.} \label{fig2}
\end{figure}

The mentioned features are also seen in the spectrum of another PC
in Fig.\,3. Note, that this curve is measured in a magnetic field
of 4\,T, which have to suppress additional features on
d$^2V$/d$I^2(V)$, which could arise from SC weak links or degraded
SC regions. It is also worth to be noted, that d$V$/d$I(V)$ of
this contact increases above 30 mV, indicating a direct metallic
contact. Absence of a barrier layer at the interface is confirmed
by proximity induced superconductivity resulting in a sharp dip at
$V$=0 on d$V$/d$I(V)$ (see Fig.3, inset). Therefore, it is also
reasonably to suggest for this contact that the barrier parameter
$Z$, obtained from the fit (see Fig.\,3, inset), is caused only by
the mismatch of the Fermi velocities between the two electrodes.
According to \cite{BTK82} $Z=(1-r)/2r^{1/2}$, where r is the ratio
of the Fermi velocities. Using the fitting parameter $Z$=0.5 for
this contact we find the Fermi velocity of MgB$_2$ about 5$\times
10^7$ cm/s taking 1.4$\times 10^8$ cm/s as the Fermi velocity of
Ag.

\begin{figure}
\includegraphics[width=\columnwidth]{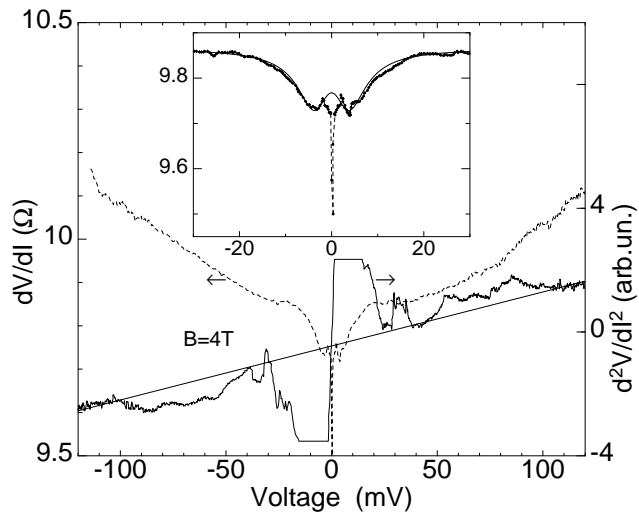}
\caption[] {First d$V$/d$I(V)$ (dashed) and second
d$^2V$/d$I^2(V)$ (solid) derivative of $I-V$ characteristics for
MgB$_2$-Ag contact at 4.2 K and magnetic field 4 T. The
d$^2V$/d$I^2(V)$ curve is truncated near zero bias due to large
signal caused by the SC gap. Straight line shows an approximate
background corresponding to the parabolic shape of the first
derivative. Inset demonstrates fitting of d$V$/d$I(V)$ (symbols
connected by dashed line) by Eq.(1) (thin solid line). Parameters
of the fit are: $\Delta$=3.8 meV, Z=0.5, $\Gamma$=3 meV. Dip in
d$V$/d$I(V)$ at $V$=0 is tentatively due to the proximity effect.}
\label{fig3}
\end{figure}

After subtracting a linear background, as shown in Fig.\,3, the
presented in Fig.\,4 PC EPI spectrum (curves 1,2), has above 30
meV similarities with the phonon DOS measured by neutron
scattering \cite{Yildirim,Osborn}. Below 30 meV all the spectra
(see Fig.2,3) exhibit a steep increase connected with
superconductivity (Andreev reflection). We have tentatively
subtracted the nonlinear background in this region, as shown in
inset of Fig.4, and found an intensive peak at 20 meV and around
30 meV (Fig.4). In principle, these peaks (as the other
lower-frequency peaks not revealed yet) might not related directly
to the EPI function, and could be due, e.g., to the suppression of
the SC order parameter. This suppression can occur near the
characteristic phonon energies with small group velocity,
corresponding to peaks in phonon DOS, as observed for PC spectra
of ordinary superconductor tantalum \cite{Yanson87}. Indeed,
corresponding peaks may be found in the mentioned energy region at
16-17, 24, 31 meV \cite{Mura,Sato} or hillock at 20 meV
\cite{Yildirim} in the phonon DOS determined by neutron
scattering. Interesting, that PC EPI spectrum of pure Mg has
maxima at 17 and 28 meV (see Fig.4a)

From the comparison of our data with the phonon DOS follows that
all phonons above 30 meV contribute in average with approximately
equal weight to the PC EPI function. We do not see the prevailing
of one mode around 60-70 meV, as follows from calculated
Eliashberg or transport EPI function in
Ref.\cite{Kong,Bohnen,Liu}. Probably, the reason is that our
measurements are mainly along the c-axis.

Concerning the estimation of an absolute value of EPI function or
EPI constant $\lambda$ from PC data, it is prematurely because the
nonlinearity of PC $I-V$ curves in the region of characteristic
phonon frequencies is too low. As is seen from Fig.3, d$V$/d$I$
increases above 30 meV only on a few percent contrary to 10-50\%
increase in the zero bias region due to the Andreev reflection at
zero magnetic field. The short elastic electron mean free path in
the constriction (diffusive regime) is likely the main reason of
the small nonlinearity. Sure, to receive both quantitative results
and final shape of EPI function the investigations of more perfect
samples or even single crystals are very desirable.

\begin{figure}
\includegraphics[width=\columnwidth]{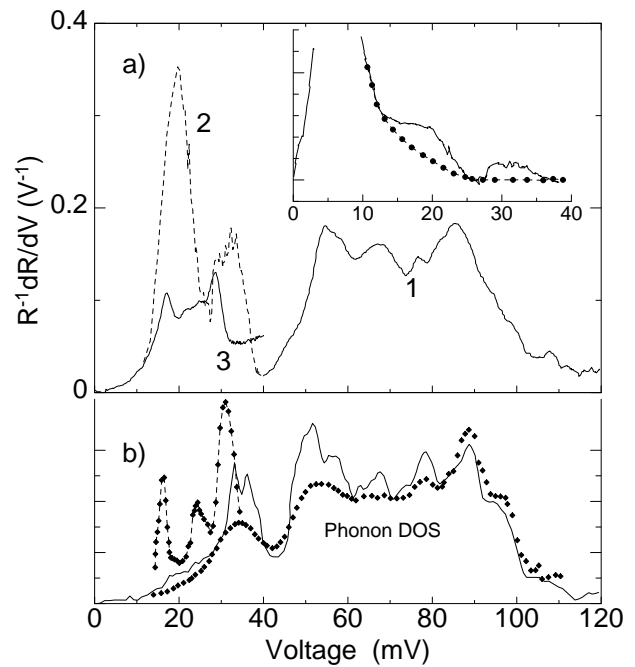}
\caption[] {a) PC EPI function (see Eq.(2)) of MgB$_2$ (curves 1
and 2) reconstructed from d$^2V$/d$I^2(V)$ in Fig.3 averaging for
plus and minus polarity after subtracting linear (as is shown in
Fig.3) and hand-made (see inset) background. Solid curve (3) shows
PC spectrum of pure Mg \cite{Naid81Mg}. Inset: proposed nonlinear
background (symbols) below 30 meV for the same contact recorded
with enlarged scale. b) Phonon DOS according neutron measurements
\cite{Yildirim} (solid line), \cite{Osborn} (symbols) and
\cite{Mura} (symbols connected by dashed line). In the latter case
only low energy ($<$40meV) part of phonon DOS is shown. }
\label{fig4}
\end{figure}

\section{Conclusion}
Our results on MgB$_2$ c-axis oriented thin film investigated by
point contacts unequivocally indicate the presence of two SC gap
with the theoretically predicted ratio 1:3 \cite{Liu}. Above SC
gap the EPI features dominate in the PC spectra. Thus the
electron-phonon coupling in MgB$_{2}$ must therefore be included
in any microscopic theory of superconductivity. The PC EPI
function has similarities above 30 meV with the phonon DOS, while
maxima at 20 and 30 meV are also resolved pointing that these low
frequency phonon modes should be considered by describing of SC
state of MgB$_2$. Using the barrierless junction, the Fermi
velocity 5$\times 10^7$ cm/s in MgB$_2$ have been estimated.

\begin{acknowledgments}
The work in Ukraine was supported by the State Foundation of
Fundamental Research under Grant $\Phi$7/528-01.  The work at
Postech was supported by the Ministry of Science and Technology of
Korea through the Creative Research Initiative Program.
\end{acknowledgments}


\begin{thebibliography}{}
\bibitem{rho}There is a scattering in $\rho_0$ for the similar films between
different publications \cite{Tu,Lee01,Lee02}.
\end{thebibliography}
\end{document}